\documentclass[aps,prl,twocolumn]{revtex4}
\usepackage{graphicx}

\newcommand{\Eb}{\mbox{\bf E}}
\newcommand{\Jb}{\mbox{\bf J}}
\newcommand{\ub}{\mbox{\bf u}}

\newcommand{\Ub}{\mbox{\bf U}}

\newcommand{\del}{\mbox{\boldmath{$\nabla$}}}
\newcommand{\nhat}{\hat{n}}

\begin{document}

\title{ Induced-charge Electrokinetic Phenomena:\\
Theory and Microfluidic Applications }

\author{Martin Z. Bazant} 
\email{bazant@mit.edu}
\affiliation{Department of Mathematics and Institute for Soldier
Nanotechnologies \\ Massachusetts Institute of Technology, Cambridge, MA
02139}
\author{Todd M. Squires}
\affiliation{Departments of Applied and Computational Mathematics
and Physics\\ California Institute of Technology, Pasadena, CA 91125
}

% MZB: began 7-15-02 iceo.tex
% MZB: start again from scratch 3-3-03 in Texas, finished 6/8/03
% TMS: 6/9/03 renamed iceoprl.tex, revised
% MZB: revised 6/11/03
% TMS: 6/11/03 revised, renamed iceoprl2.tex
% MZB: 6/12/03 final draft, renamed iceoprl.tex again
% TMS: 6/12/03 last final revision, renamed iceoprl5532a.tex
% TMS: 6/12/03 relented and changed name back to iceoprl.tex
% MZB: 6/13/03 thought about new name, and submitted as IcEoPrL.tex
% MZB: 9/23/03: copied to iceoprl2.tex, revision on 10/14/03
% MZB: 2/2/04: copied to iceoprl3.tex, tiny changes for final version

\date{\today}

\begin{abstract}

We give a general, physical description of ``induced-charge
electro-osmosis'' (ICEO), the nonlinear electrokinetic slip at a
polarizable surface, in the context of some new techniques for
microfluidic pumping and mixing. ICEO generalizes ``AC
electro-osmosis'' at micro-electrode arrays to various dielectric and
conducting structures in weak DC or AC electric fields. The basic
effect produces micro-vortices to enhance mixing in microfluidic
devices, while various broken symmetries --- controlled potential,
irregular shape, non-uniform surface properties, and field gradients
--- can be exploited to produce streaming flows. Although we emphasize
the qualitative picture of ICEO, we also briefly describe the 
mathematical theory (for thin double layers and weak fields) and apply
it to a metal cylinder with a dielectric coating in a suddenly applied
DC field.
\end{abstract}

\maketitle

The advent of microfluidic technology raises the fundamental
question of how to pump and mix fluids at micron scales, where
pressure-driven flows and inertial instabilities are suppressed by
viscosity~\cite{stone01,beebe02}. The most popular
non-mechanical pumping strategy is based on
electro-osmosis --- the effective slip, $\ub_\|$, at a
liquid-electrolyte/solid interface due to tangential electric field,
$\Eb_\|$. The Helmholtz-Smoluchowski formula,
\begin{equation}
\ub_\| = - \frac{\varepsilon \zeta}{\eta} \, \Eb_\| , \label{eq:slip}
\end{equation}
gives the slip in terms of the permittivity, $\varepsilon$, and
viscosity, $\eta$, of the liquid and the zeta potential, $\zeta$,
across the diffuse part of the (thin) interfacial double
layer~\cite{lyklema}. The usual
%``fixed-charge''
case of constant (possibly non-uniform~\cite{ajdari95}) $\zeta$,
however, has some drawbacks related to linearity, $\ub_\| \propto
\Eb_\|$: (i) the flow is somewhat weak, e.g. $u_\|=70 \mu$m/s in
aqueous solution with $E_\| =100$ V/cm and $\zeta=10$ mV and (ii) AC
fields, which reduce undesirable Faradaic reactions, produce zero
time-averaged flow.

These drawbacks do not apply to AC electro-osmosis, recently
discovered by Ramos et al.~\cite{ramos98} and
Ajdari~\cite{ajdari00}. Nonlinear electro-osmotic slip is produced at
micro-electrodes as an AC field acts on {\it induced} double-layer
charge prior to complete screening. In spite of extensive work,
however, this promising effect remains limited to quasi-planar
pairs~\cite{aceo} or arrays~\cite{acpumping} of electrodes at a single
AC frequency, $\omega_c = \tau_c^{-1}$, where $\tau_c = \lambda_D L/D$
is the ``RC time'' of an equivalent circuit of bulk resistors of size
$L$ (the electrode spacing) and double-layer capacitors of thickness,
$\lambda_D$, the Debye screening length, and $D$ is an ionic
diffusivity.

How general is this phenomenon?  Nonlinear electro-osmotic flows have also
been observed at dielectric impurities on electrodes with AC
forcing~\cite{nadal02b} and, more suggestively, at dielectric
(non-electrode) micro-channel corners in DC
fields~\cite{thamida02}. Although it is largely unknown (and uncited)
in the West, similar flows have also been studied in the Russian
literature on polarizable colloids~\cite{murtsovkin96}, including the
effect of such flows on dielectrophoresis~\cite{simonova01}. The
unifying principle in these diverse effects is that an applied field
acts on its own induced diffuse charge, so we suggest the term,
``induced-charge electro-osmosis'' (ICEO), to describe it.

In this Letter, we describe more general ICEO flows which may be
produced in microfluidic devices.  Before we begin, we stress that
ICEO dominates electrokinetics at small total zeta potentials (initial
+ induced), $\zeta \ll 2(kT/e)\log(L/\lambda_D)$, where
$kT/e \approx 20$ mV is the thermal voltage. When this
condition is violated, as in highly charged colloids, surface
conduction leads to other ``non-equilibrium electrosurface phenomena''
(NESP)~\cite{dukhin93},
% which may influence in colloidal
%aggregation near electrodes~\cite{trau97,yeh97,sides01,nadal02a}.
and at very large voltages (above a limiting current) bulk space
charge may also drive ``second-kind electrokinetic
phenomena''~\cite{dukhin91}. Such exotic effects may be useful
in microfluidics~\cite{ben02}, but we focus on small voltages,
which are often preferable in real devices.

\begin{figure*}
\includegraphics[width=1.5in]{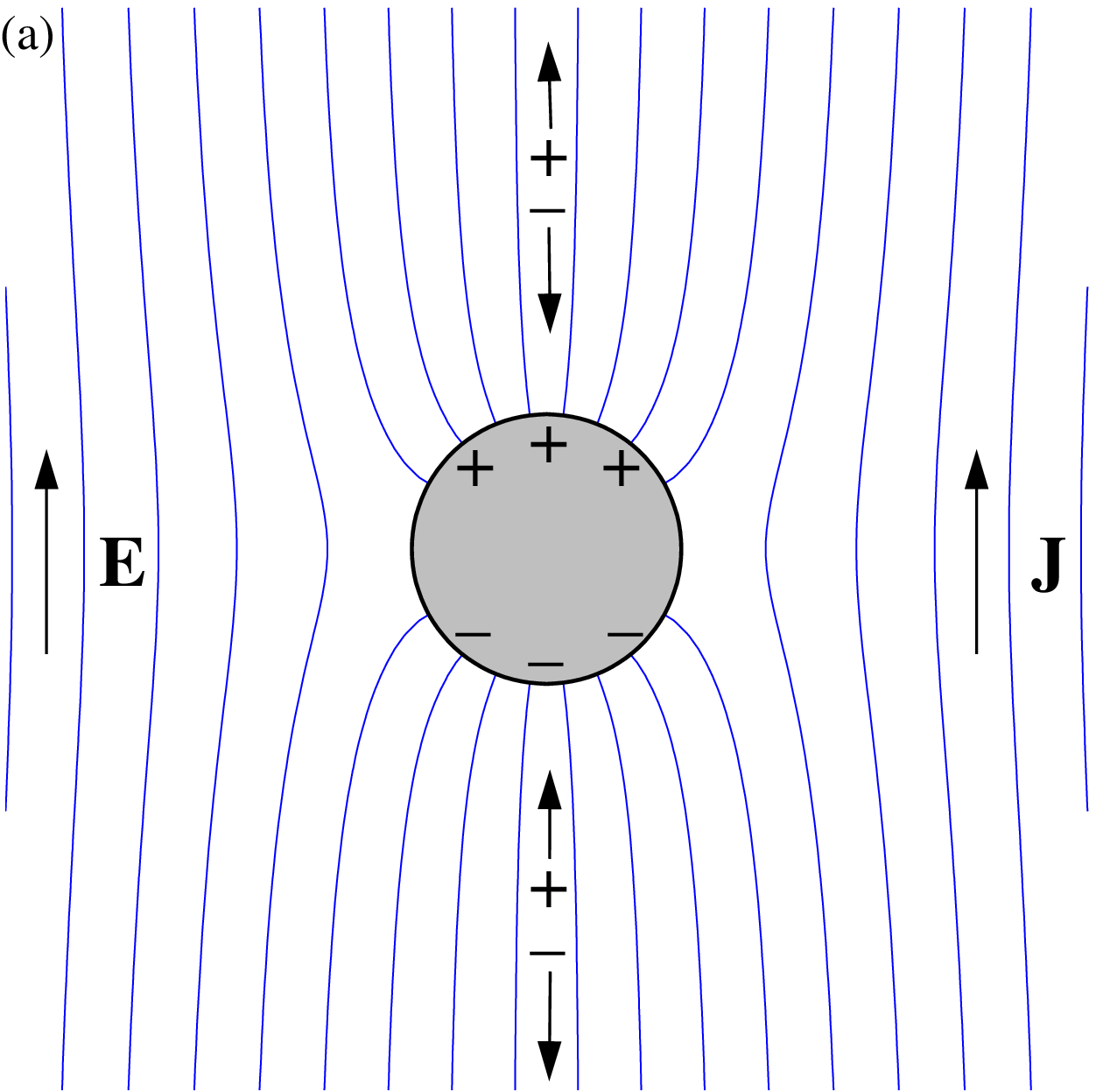}\hspace{0.1in}  \nolinebreak
\includegraphics[width=1.5in]{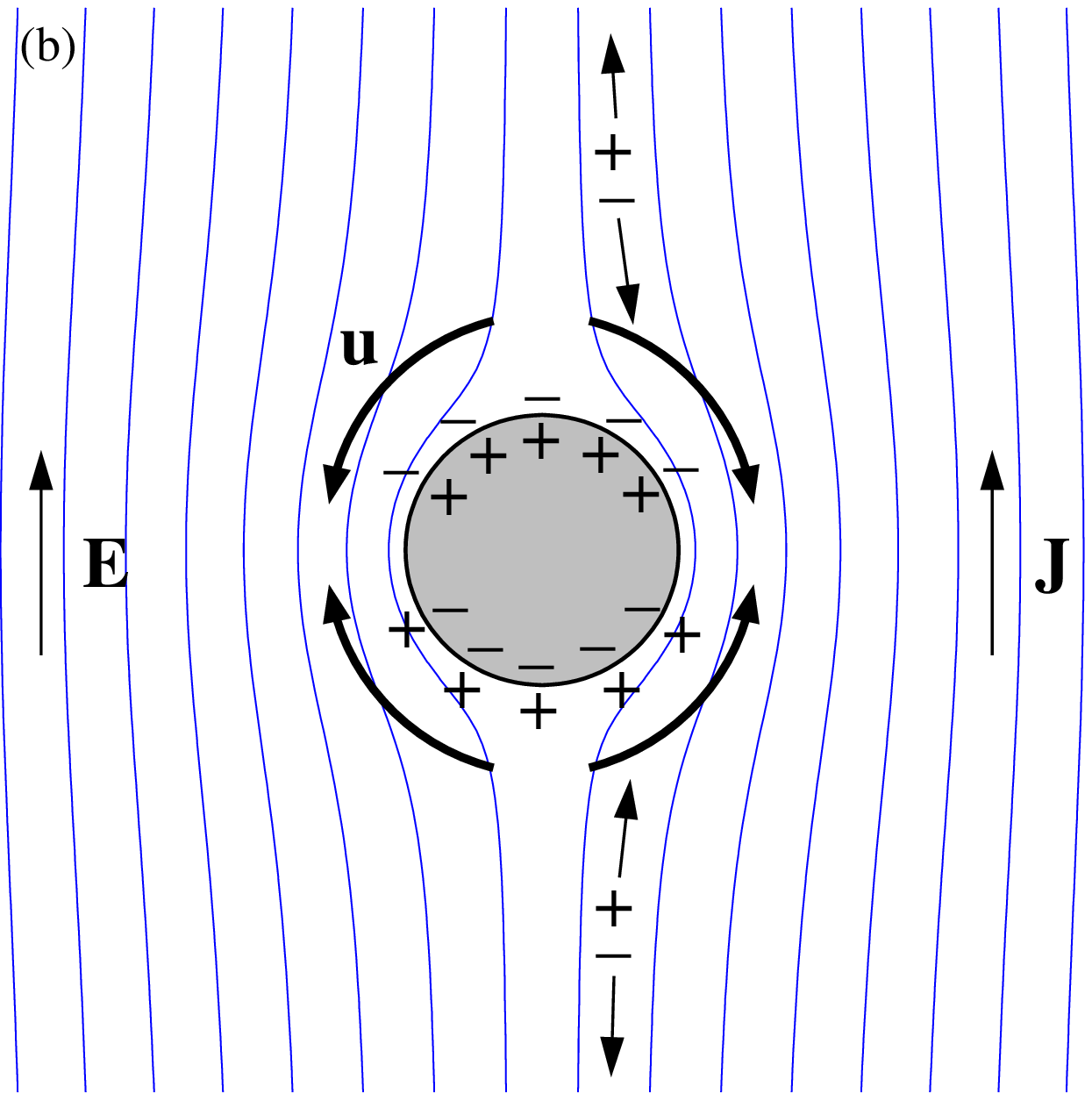}\hspace{0.1in}  \nolinebreak
\includegraphics[width=1.5in]{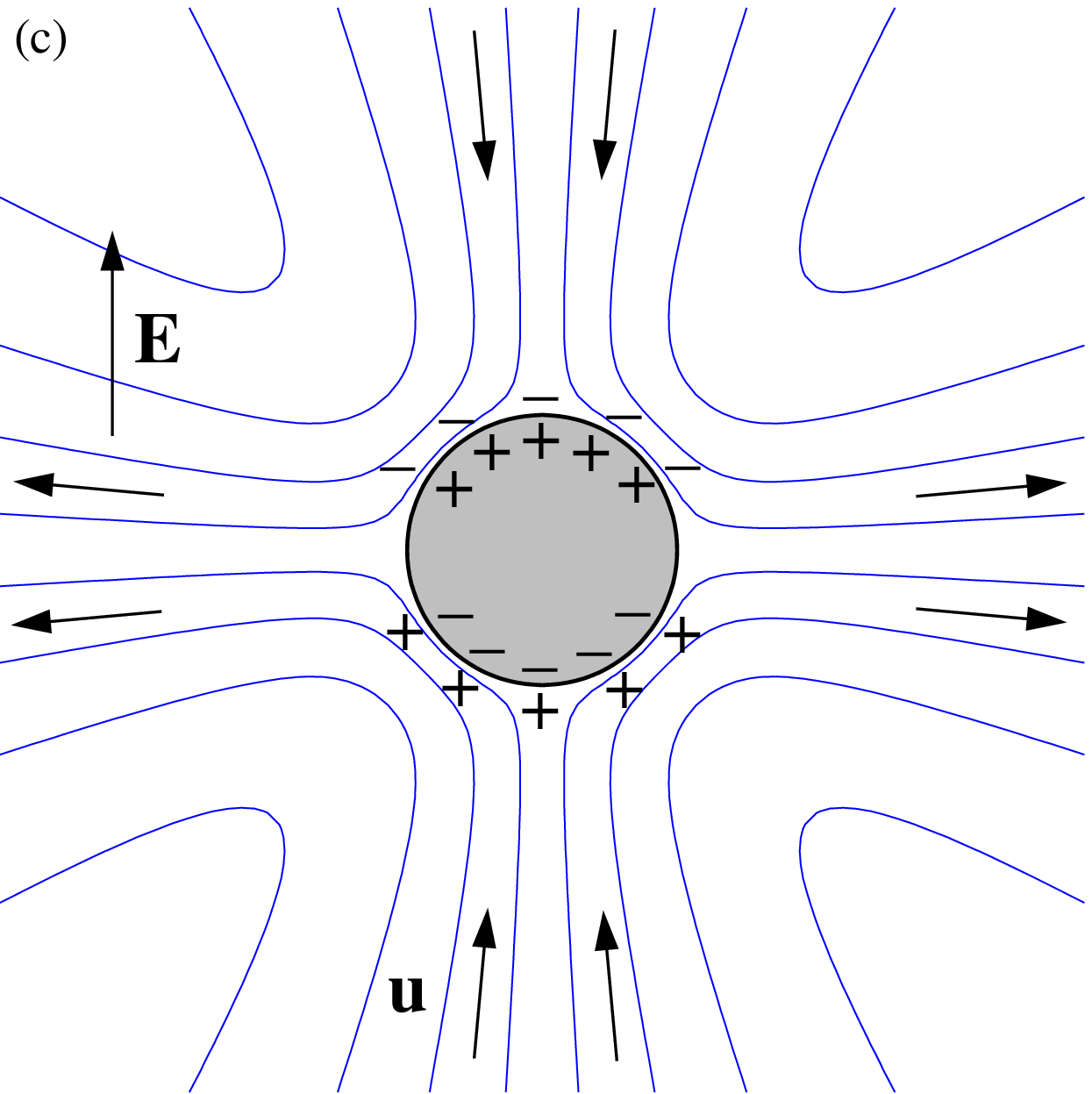}\hspace{0.1in} \nolinebreak
\includegraphics[width=1.5in]{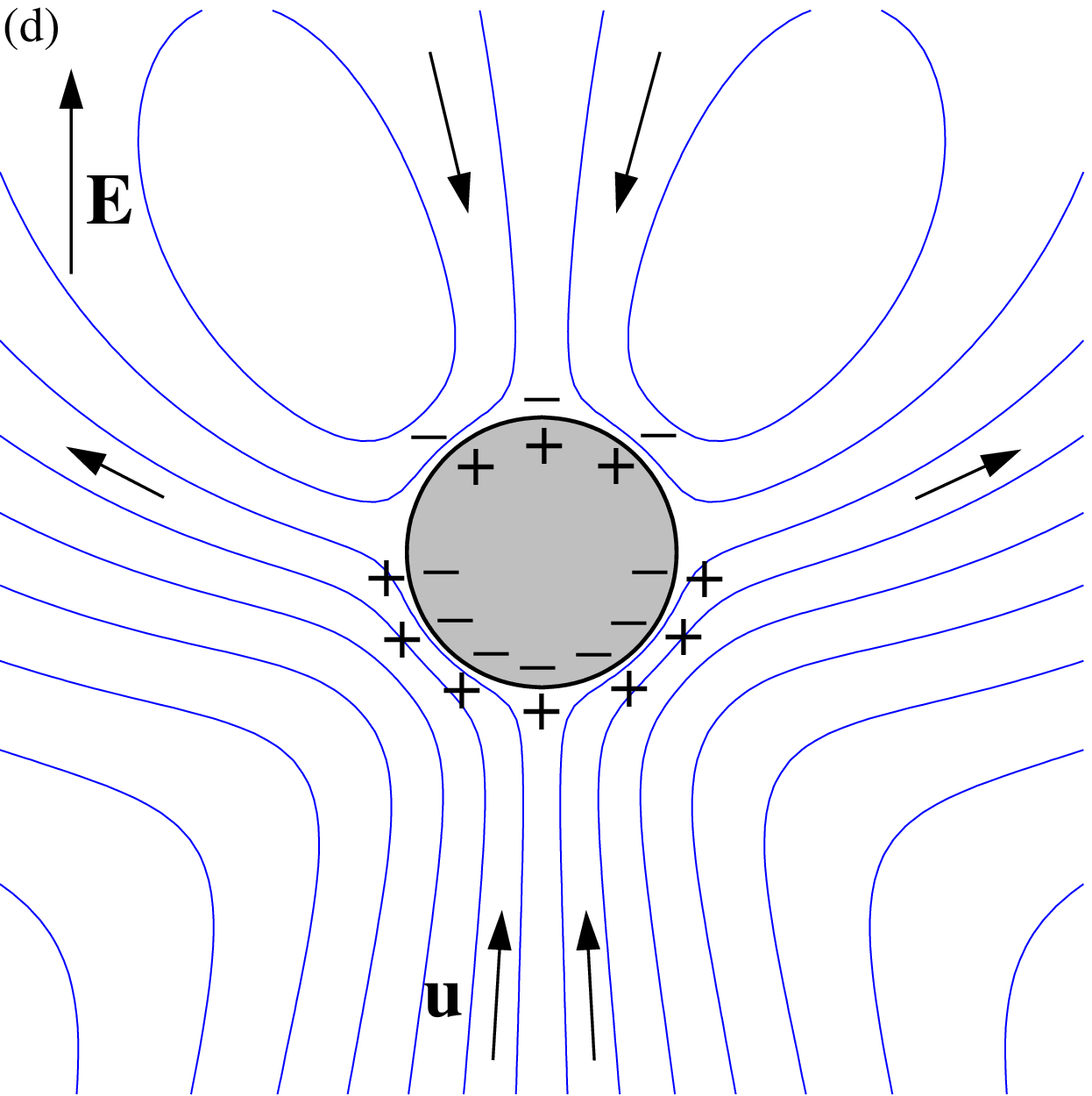}
\caption{\label{fig:cylinder} Lines of electric field $\Eb$ (or current,
$\Jb=\sigma\Eb$) around a cylindrical metal wire in an electrolyte (a)
before and (b) after double-layer charging in response to a suddenly
applied DC field
% The electronic charge on the sphere
% and net ionic charge in the electrolyte are also sketched.
and (c) the resulting ICEO streamlines. The flow around a
charged polarizable cylinder is shown in (d).
% , in a frame moving with the
% electrophoretic velocity (which is the same for a charged
% non-polarizable sphere).
 }
\end{figure*}

%{\it Physical Picture.}

ICEO is nicely illustrated by the flow around an uncharged, ideally
polarizable cylinder (e.g. an inert metal wire) of radius $a$ in a
suddenly applied, weak, uniform field, $\Eb_b$. (The analogous problem
for a metal colloidal sphere was introduced long ago by
Levich~\cite{levich}; the double-layer polarization was later analyzed
in more detail by Simonov and Shilov~\cite{simonov73b} and the
resulting flow by Gamayunov, Murtsovkin and
A. S. Dukhin~\cite{gama86}.) As shown in Fig.~\ref{fig:cylinder}(a),
the initial electric field lines
%(after very fast electronic relaxation)
are those of a conducting sphere in vacuum, perpendicular to the
surface, but the ionic current, $\Jb = \sigma \Eb$, affects the field.
Without surface conduction or Faradaic reactions to transfer the
normal current at the surface, ionic charge accumulates in the
double-layer ``capacitor''. As shown in Fig.~\ref{fig:cylinder}(b), a
steady state is reached when the bulk field lines are expelled to
become those of an insulator, parallel to the surface.
% By the same argument,
% this actually holds for any inert object, regardless of its electrical
% properties or shape~\cite{levich}.
The induced diffuse charge (or $\zeta$)
shown in Fig.~\ref{fig:cylinder}(b) is non-uniform --- negative
($\zeta>0$) where the initial current leaves the surface and positive
($\zeta<0$) where it enters. 
% By
% symmetry, $E_\|=0$ at the poles ($\theta=0,\pi$) and $\zeta=0$ at the
% equator ($\theta=\pi/2$), so nonzero slip occurs only at intermediate
% angles
Since $\ub_\| \propto - \zeta \, \Eb_\|$, we can anticipate the
quadrupolar ICEO flow shown in Fig.~\ref{fig:cylinder}(c), which draws
in fluid along the field axis and ejects it radially. 
%
% By symmetry,
%the flow cannot depend on the sign of $\Eb_b$, so it persists in both
%DC and AC fields (up to the charging frequency, $\omega_c =
%\frac{D}{\lambda_D a}$). Indeed, reversing $\Eb$ changes the sign of
%the induced $\zeta$ everywhere, producing the same slip velocity,
%$\ub_\|$.
%
Reversing $\Eb$ changes the sign of the induced $\zeta$ everywhere, so
the same flow also persists in AC fields (up to the charging
frequency, $\omega_c = D/\lambda_D a$).

The magnitude of the flow follows easily from dimensional
analysis. After charging,
% ($t \gg \tau_c$),
the double layer acquires the background voltage across the object
(non-uniformly), which produces zeta potentials of order $E_b a$, and
from Eq.~(\ref{eq:slip}) typical flow speeds of
order,  
\begin{equation}
U_0 =  \frac{\varepsilon a E_b^2}{\eta}  \label{eq:Uo}
\end{equation}
(as for a colloidal sphere~\cite{gama86}).  When the applied voltage,
$E_b a$, exceeds typical equilibrium zeta potentials (10mV), ICEO flow
exceeds that of standard DC electro-osmosis, e.g. $U_0 = 0.7$mm/s in
water for $E_b =100$ V/cm and $a=10\mu$m. In that case, the maximum
frequency, $\omega_c$, is of order 10 kHz for $\lambda_D=10$nm. (For
these parameters, we are also justified in neglecting surface
conduction~\cite{lyklema,dukhin93}.)

\begin{figure}
\includegraphics[width=1.5in]{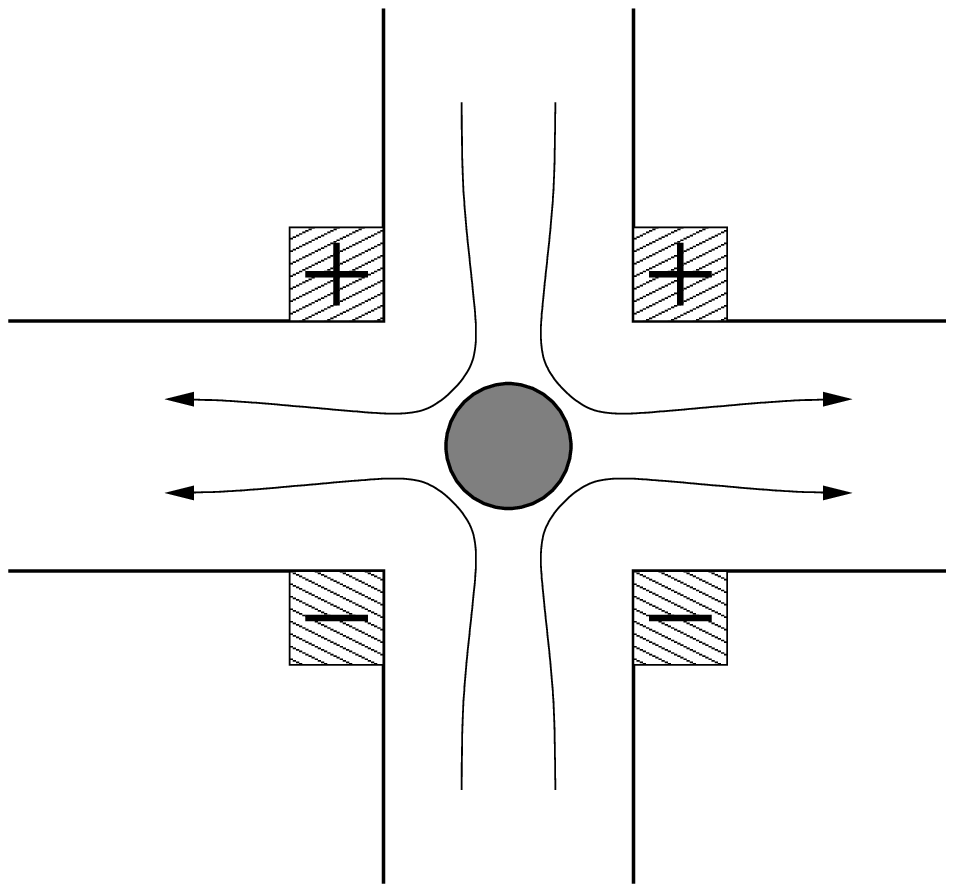}
\hspace{0.2in}\nolinebreak
\includegraphics[width=1in]{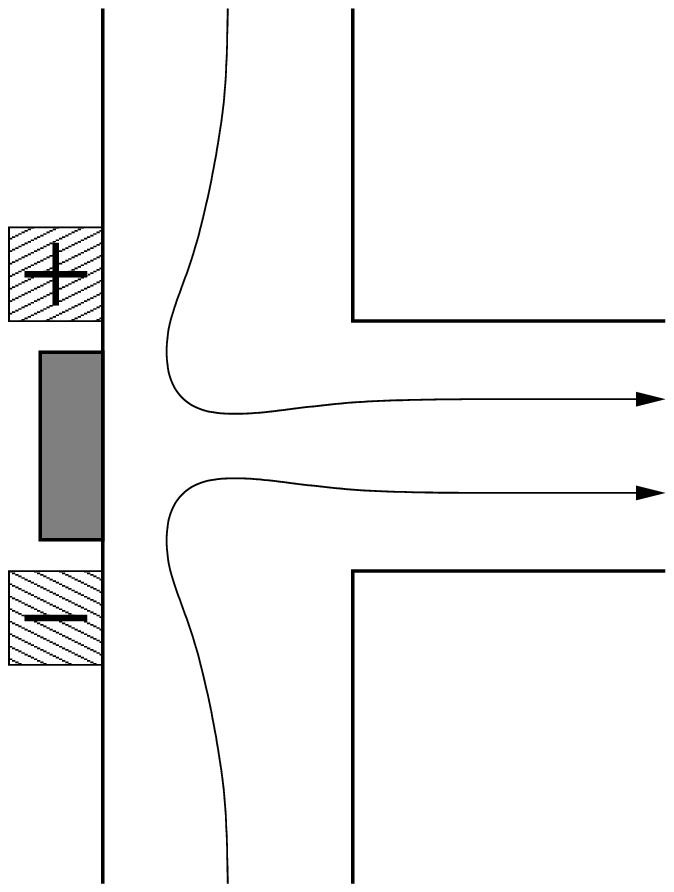}
\\
\vspace{0.1in}
\includegraphics[width=2in]{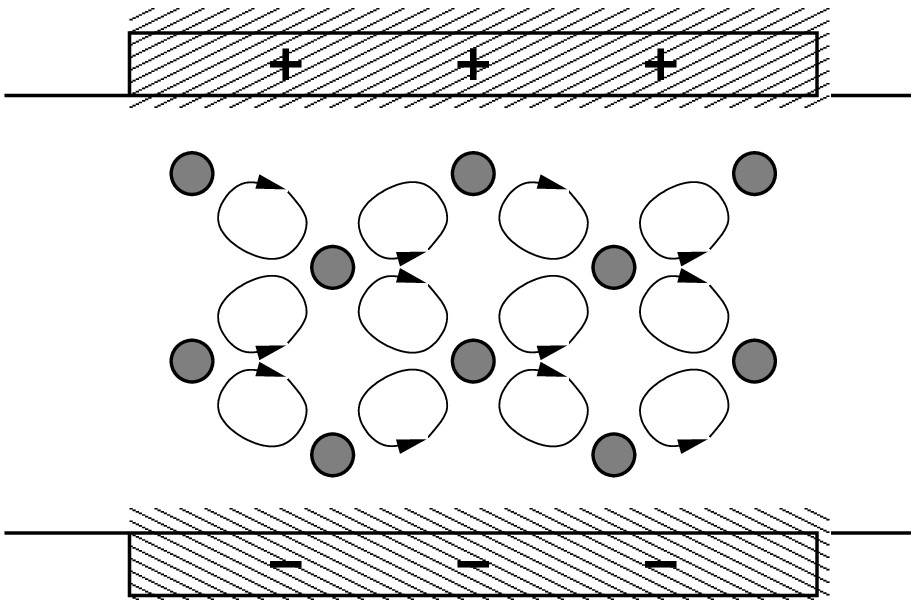}
\caption{\label{fig:symdevices} Simple microfluidic devices exploiting
ICEO flows around fixed metal objects (shaded) driven by small AC
voltages at micro-electrodes (cross-shaded).  }
\end{figure}

The symmetry of ICEO flow in Fig.~\ref{fig:cylinder}(c) suggests the
microfluidic devices sketched in Fig. ~\ref{fig:symdevices}. For
example, a metal post in a transverse AC or DC field produces local
time-averaged flow as in Fig.~\ref{fig:cylinder}(c).  Placed at a
cross junction with two pairs of corner electrodes
(Fig. ~\ref{fig:symdevices}a), ICEO draws liquid in along one channel
and forces it out the other, and the flow is easily reversed by
changing the field direction. An array of posts in a transverse AC
field (Fig. ~\ref{fig:symdevices}b) produces micro-vortices to enhance
mixing in passing flows. A different design for a T-junction pump
(above right) employs a metal surface coating, which could wrap around
in the third dimension to reduce viscous drag by replacing more of the
channel wall with sources of ICEO. (See also
Fig.~\ref{fig:asymdevices}.)

%Surface-generated pressures of order $\eta U_0/a = \varepsilon E_0^2$
%decay like $(a/r)^2$, so a single post in a channel of width $L$
%produces a pressure head of order mPa.  This suggests that simple ICEO
%devices are best suited for local manipulation of fluids.  Stronger
%flows may be achieved with other designs below, where some of the
%channel surface (causing viscous drag) is replaced by polarizable
%elements (causing flow), and streaming flows require broken symmetry.

%{\it Variable charge and potential. }

In such devices, streaming flows are easily produced by broken
symmetries.  As the first example, consider a metal cylinder of
non-zero total charge.  In a DC field, the ICEO flow shown in
Fig.~\ref{fig:cylinder}(d) (and given below) is simply a superposition
of the nonlinear quadrupolar flow described above and the linear
streaming flow of electrophoresis, which averages to zero for a freely
suspended object. In microfluidic devices, however, there is a new
possibility: By controlling an object's potential, its {\it induced
total charge} can be made to vary in phase with the applied field to
produce time-averaged AC streaming. For example, a thin metal post at
position, $x_0$, and potential, $\phi_0$, between two electrodes
imposing a linear potential, $\phi = -E_b x = Vx/L$, generates a
streaming flow of order,
\begin{equation}
U_1 = \frac{\varepsilon(\phi_0 + E_b x_0) E_b}{\eta} .
% = \frac{\varepsilon}{\eta}\left(x_0 - \frac{\phi_0L}{V}\right)
% \left(\frac{V}{L}\right)^2 .
% = \frac{\varepsilon V (V x_0 - \phi_0 L)}{\eta L^2}
% =\left(\frac{x_0}{L} - \frac{\phi_0}{V}\right)\, U_0
\label{eq:U1}
\end{equation}
In the case of the mixer in Fig.~\ref{fig:symdevices}c, a post
grounded to an electrode, $\phi_0=V$, pumps toward the nearest wall
with speeds larger than the (superimposed) fixed-total-charge flow by
a geometry-dependent factor, $|U_1/U_0| = (L-x_0)/a$. Since posts of
prescribed potential act as electrodes, this manifestation of ICEO
resembles AC electro-osmosis~\cite{ramos98,ajdari00}, but it does not
require AC background fields. It is perhaps closer to DC
``field-effect flow control''~\cite{schasfoort99}, although ICEO
involves metal surfaces at much smaller voltages coupled directly to
the primary electrodes.

\begin{figure}
\includegraphics[width=3in]{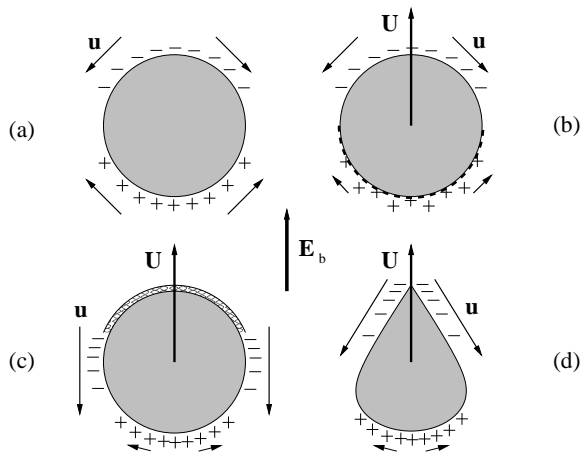}
\caption{\label{fig:asymposts} Sketches of induced diffuse charge
  ($+,-$) and ICEO slip,$\ub$, around (a) a circular cylinder, (b) with
  a partial coating (dashed) of increased surface capacitance, (c)
  with an insulating coating (shaded), and (d) an asymmetric cross
  section. In (b)-(d), the ICEP velocity, $\Ub$, points up, regardless
  of whether the applied field, $\Eb_b$, points up or down. }
\end{figure}

%{\it Asymetric Objects and Non-uniform Fields. }

As in the case of AC pumping at electrode
arrays~\cite{ajdari00,acpumping}, other broken symmetries also cause a
polarizable object to pump fluid via ICEO, if
fixed, or move via ``induced-charge electrophoresis'' (ICEP), if
freely suspended, at the velocity scale, $U_0$, in either DC or AC
fields.
%Such rectification is due to asymmetry in either double-layer charging
%or tangential-field evolution, which generally produces a larger slip
%velocity on one side of the object than on the other.
For example, consider a metal cylinder with {\it non-uniform surface
properties} in a uniform field, as in
Fig.~\ref{fig:asymposts}(b)--(c). If one side has a greater surface
capacitance as in (b), then some induced charge ends up immobilized on
the surface unable to cause slip, so the other side ``wins'' in
driving ICEP in its direction. If the sphere is partially insulating
as in (c), then primarily the conducting portion is polarized, and the
steady-state field acts on the shifted diffuse charge to cause ICEP
toward the insulating side.

If the object has an {\it irregular
shape} as in (d) ICEP occurs in the more protruding direction, where
the induced charge is better aligned with the tangential field (as in
some experiments~\cite{gama87}).  If the same object in (d) were
rotated to break left-right symmetry, it would move in the opposite
(more rounded) direction, perpendicular to the field axis, albeit
unstable to rotations restoring field alignment.  Such ``transverse
electro-osmosis'' has been demonstrated for fixed-charge surfaces in DC
fields~\cite{ajdari95}, but with polarizable surfaces it can also
occur in AC fields.

The ICEO flows in Fig.~\ref{fig:asymposts} motivate the microfluidic
devices in Fig.~\ref{fig:asymdevices}, which pump in one direction
while mixing via superimposed circulating flows in either AC or DC
fields.  For example, asymmetric posts pump and mix in response to an
applied field along the channel (a), and the pumping
direction may be reversed with a transverse field (not shown).  An
equivalent design (b) involves asymmetric grooves in a
metallic channel wall. Such features may wrap around the channel in
the third dimension, further reducing viscous drag.

\begin{figure}
\includegraphics[width=2.3in]{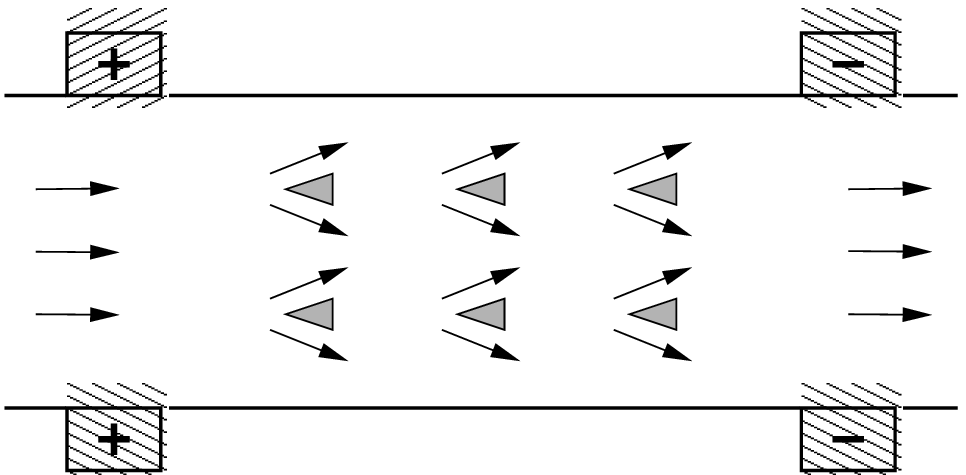}
\\
\vspace{0.1in}
\includegraphics[width=2.3in]{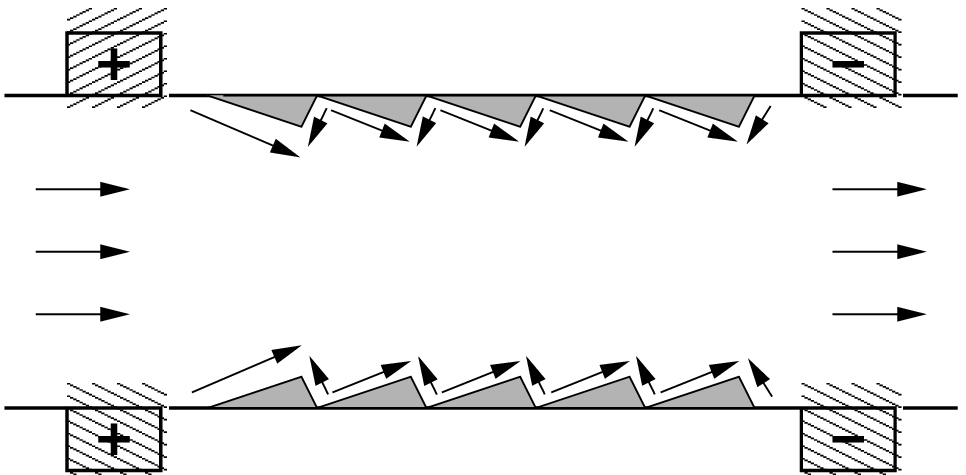}
\caption{\label{fig:asymdevices}
Microfluidic devices exploiting asymmetric ICEO.  }
\end{figure}

Finally, we consider {\it non-uniform applied fields}.  For example, an
uncharged metal post in a non-uniform DC or AC field pumps down the
field gradient, $\nabla E_b$, with a typical streaming velocity,
\begin{equation}
U_2 = \frac{\varepsilon a^2}{\eta} |\nabla E_b|^2 ,
\end{equation}
which follows from Eq.~(\ref{eq:slip}) since induced variations in
$\zeta$ are of order, $a^2 |\nabla E_b|$. Such flows profoundly
influence dielectrophoresis for polarizable colloidal
spheres~\cite{simonova01} (via ICEP), but non-spherical shapes require
further study.  More generally, non-uniform fields provide another
useful degree of freedom in ICEO devices. As a simple example, one
transverse electrode in the mixer of Fig.~\ref{fig:symdevices} could
be shortened to produce additional pumping toward the opposite wall
(and large-scale circulation) driven by the resulting field
gradient. All of this flexibility raises interesting open questions of
optimal design.

The physical arguments above can be justified by mathematical analysis
of the electrokinetic equations~\cite{lyklema} for weak fields and
thin double layers, as described in subsequent papers, beginning with
Ref.~\cite{squires03a}.  The origin of ICEO and other NESP is an
effective boundary condition on the neutral
bulk electrolyte which expresses the conservation of the diffuse
charge, $q$, in a thin double layer:
\begin{equation}
- \frac{\partial q}{\partial t} = J_F + \nhat \cdot \Jb +
  \del_s\cdot\Jb_s \label{eq:genbc}
\end{equation}
where $J_F$ the Faradaic current, $\nhat$ the outward normal, $\del_s$
the surface gradient, and $\Jb_s$ the surface current.  NESP mostly
involve steady surface conduction, $\nhat \cdot \Jb =
-\del_s\cdot\Jb_s$, at low AC frequencies, $\omega \ll D/a^2$, which
produces concentration gradients affecting electro-osmosis at highly
charged surfaces~\cite{murtsovkin96,dukhin93}.

In contrast, ICEO derives from the dominant ``$RC$ coupling'' for
small $\zeta$ (initial + induced) in a homogeneous electrolyte:
\begin{equation}
C_D \,\frac{ \partial \zeta}{\partial t} 
= \sigma\, \nhat\cdot\Eb   \label{eq:nonlinbc}
\end{equation}
where $C_D(\zeta)=-\frac{dq}{d\zeta}$ is the nonlinear differential
capacitance per unit area. (Ideal polarizability, $J_F=0$, is often a
good approximation for small $\zeta$ and AC forcing.)  The variable
$\zeta$ in Eq.~(\ref{eq:slip}) is then obtained by solving Laplace's
equation, $\nabla^2\phi=0$, subject to the boundary condition
(\ref{eq:nonlinbc}). For linear screening, $C_D =
\varepsilon/\lambda_D$, such analysis has been done for
micro-electrode arrays~\cite{ramos98,ajdari00,aceo}, while nonlinear
screening has also been discussed in colloids~\cite{simonov73b}.

Motivated by the devices in Fig.~\ref{fig:symdevices}, we calculate
the ICEO slip for a metal cylinder at potential
$\phi_0$ (or $\zeta_0 = (\phi_0-\phi_b)/(1+\delta)$) in a suddenly
applied, weak background potential $\phi \sim \phi_b - E_b\, r
\cos\theta$, where $\zeta = (\phi-\phi_0)/(1+\delta) \ll
kT/e$ at $r=a$. (Here $\delta$ is the capacitance ratio of the
compact part to the diffuse part of the double layer~\cite{ajdari00}.)
The bulk potential,
\begin{equation}
\phi(r,\theta,t) = \phi_b  - E_b \, r \cos\theta \left(1 +
g(t) \frac{a^2}{r^2}\right)  \label{eq:phisphere}
\end{equation}
involves an induced dipole moment, $g(t) = 1 - 2\, e^{- t/\tau_q}$,
effectively changing from a conductor ($g=-1$, $\tau \ll \tau_q$) to
an insulator ($g=1$, $t \gg \tau_q$) at the time scale, $\tau_q =
\tau_c/(1+\delta)$.  (For nonlinear screening, the poles charge
more slowly, $dC_D/d|\zeta|\geq 0$, but the steady state is the
same.)
The slip velocity, $\ub_\| = u_\theta \hat{\theta}$, has two
terms:
\begin{equation}
u_\theta(\theta,t) = U_1^\prime \, f(t) \, \sin\theta + 2 \,
U_0^\prime\, f(t)^2 \, \sin 2\theta .  \label{eq:utheta}
\end{equation}
where $f(t) = 1+g(t)$, $U_1^\prime = \varepsilon (\phi_b-\phi_0)
E_b/\eta(1+\delta)$, and $U_0^\prime = U_0/(1+\delta)$.  The
first term produces pumping past the cylinder 
%(or electrophoresis if
% it is not fixed), 
while the second, peaked at $45^\circ$ to the field and growing more
slowly, produces a symmetric quadrupolar flow. Both terms have have
non-zero time averages in AC fields, $\langle u_\theta^{(1)} \rangle
\propto \langle \zeta_0 E_b \rangle \sin\theta$ and $\langle
u_\theta^{(0)} \rangle \propto a \langle E_b^2 \rangle \sin 2\theta$,
respectively, although the former requires controlling
$\phi_0$. Finally, we note that a thin dielectric layer (modeled by
$\delta>0$ as in Ref.~\cite{ajdari00}) reduces ICEO flow by
$1/(1+\delta)$, so care must be taken keep metal surfaces clean
in real devices.

In summary, ICEO is a versatile technique for microfluidic pumping
using weak AC (or DC) fields, including many new possibilities other
than AC electro-osmosis at planar micro-electrode arrays.  The
remarkable richness of ICEO flows merits further analysis and
experiments, especially at moderate voltages where other NESP may
become important. Nevertheless, simple drawings, as in Figs. 1-4,
suffice for a qualitative understanding in many cases.

\ 

This research was supported in
part by the U.S. Army through the Institute for Soldier
Nanotechnologies, under Contract DAAD-19-02-0002 with the U.S. Army
Research Office.

\end{document}